\documentstyle[seceq,supplement,epsf,float,wrapft,mbf]{ptptex}

\newcommand{\Tr}{{\mbox{Tr}}}
\newcommand{\bzero}{{\bf 0}}
\newcommand{\bx}{{\mbf x}}
\newcommand{\by}{{\mbf y}}
\newcommand{\br}{{\mbf r}}
\newcommand{\bR}{{\mbf R}}

\title{Quarkonium in Heavy Ion Collisions}
\subtitle{high-energy multiple scattering of quark pair in nuclei}

\author{H.~{\sc Fujii}\footnote{E-mail: hfujii@phys.c.u-tokyo.ac.jp}}

\inst{
Institute of Physics, University of Tokyo,\\
Komaba 3-8-1, Meguro, Tokyo 153-8902
}

\recdate{%      %Editorial Office will fill in this.
\today
}

\abst{
Quarkonium suppression in heavy ion collisions is 
a potential signature of the formation of the quark-gluon
plasma. After a very brief review of the J/$\psi$ result at CERN, 
we restrict our discussion to the effects of the
high-energy multiple scattering of the quark pair in the colliding
nuclei.
}

\begin{document}

\maketitle

\section{Quarkonium in heavy ion collisions}

Investigation of the QCD plasma, which is a system of the strongly
interacting gas of the quarks and the gluons,
is one of the fundamental subjects in the physics
of the strong interaction.\cite{QM01,QM02,matsui,Phenix} 
In the experimental study with
high-energy nucleus-nucleus collisions,
the J/$\psi$ production is considered as a special
signal sensitive to the plasma which is created in the early stage of
the reactions.
The color attraction between
the heavy $c\bar c$ pair produced in the initial hard
parton scatterings will be screened in the plasma, which
will suppress the binding to the physical J/$\psi$.\cite{MS86}

This predicted suppression was observed in the NA38 experiment.\cite{NA38}
However it was seen even for the proton-nucleus (p$A$) interactions
as well as the nucleus-nucleus ($AB$).
Because the exponential scaling with the effective target thickness
 $L$\footnote{One should be careful that $L$ is defined by a model.}
 is found\cite{GH92} and
the J/$\psi$ and the $\psi'$ are suppressed in the same 
rate,~\cite{NA38plb466}
this suppression is interpreted as a normal nuclear effect
or absorption of the produced
pre-resonance state by the nucleons in the nucleus with 
$\sigma_{\rm abs}=6\sim 7$ mb.\cite{GH92,KLNS97}
The large cross section and
the state independence are explained as the importance of the
color-octet pre-meson state in the production of the quarkonia.\cite{KS96}
This is a typical example to show the necessity
of the joint studies of the quarkonium 
in p$\bar{\rm p}$, p$A$ and $AB$ reactions.

The Pb-Pb collisions brought us a real surprise; 
much stronger suppression was reported.\cite{NA50}
Threshold behavior is seen in the data,\cite{NA50,QM02}
which seems very hard to explain theoretically
(even with the plasma\cite{BDO00}).
All medium effects must be studied duly and critically
before the final conclusion is reached,
although the plasma is the most immediate possibility.

The absorption of the J/$\psi$ state by the comoving secondaries
produced in the reaction is examined extensively\cite{CFK00}.
As a small color singlet object, J/$\psi$'s interactions with the
light secondaries are expected to be weak.
Quantitatively, however,
it is largely unknown with little experimental information. 
Model calculations are done for the processes
like $\psi \pi, \psi\rho \to D\bar D, D^* \bar D^*$, etc.
But the resulting cross sections at the threshold region are very
model-dependent.\cite{cross} We note that the possible modifications
of the $D$ state in medium also affect the quarkonium
dissociation rate.
\cite{STST00,AH00}

NA50 collaboration recently reported\cite{QM02} the smaller cross section
$\sigma_{\rm abs}=4.4\pm 0.5$ mb with the improved statistics,
and the J/$\psi$ yield in the peripheral region with the smaller errors,
which apparently makes the threshold clearer.
Any suppression model proposed so far
should be subjected to this data.

Strictly speaking,  the formation 
amplitude of the J/$\psi$ must be calculated
in the $AB$ collisions, where
the cross section of the J/$\psi$ or the pre-meson state with a nucleon
is unlikely to have a definite meaning. 
In \S2 we study a model in which the
nuclear effect modifies the $c\bar c$ state before forming the J/$\psi$.

At the collider energies,
the situation will change at least in two respects.
One cannot think of the independent scatterings
of the J/$\psi$ off the valence nucleons because they are
localized in the extremely thin regions in the CM frame (see \S3).
Multiple $c\bar c$-pair production may alter the assumption
of no accidental coalescence, which seems valid 
at the CERN-SPS energy owing to the rare $c\bar c$ production.\cite{coalesce}

\section{A multiple scattering model for J/$\psi$ suppression}

\begin{wrapfigure}[12]{R}{0.36\textwidth}
\epsfxsize=0.38\textwidth
\epsffile{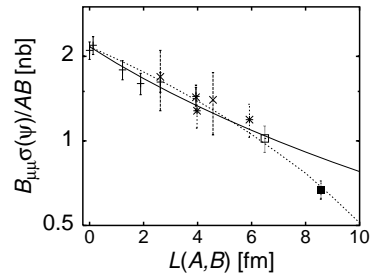}
\caption{J/$\psi$ cross section to $\mu^+\mu^-$ 
in the p$A$ and $AB$ collisions~\cite{NA50}
compared with the models with (\ref{eq:shift}) [dashed]
and with (\ref{eq:smear}) [solid], respectively.
The filled square is the Pb-Pb point.
\label{fig1}}
\end{wrapfigure}

Recently Qiu, Vary and Zhang (QVZ) proposed a new nuclear mechanism
for the quarkonium suppression, which successfully explains
the observed data at CERN\cite{QVZ02}.
In their model the inclusive cross section of 
the J/$\psi$ production in the collision of the hadrons $A$ and $B$
is written in a factorized form:
\begin{eqnarray}
&&\sigma_{AB\to {\rm J}/\psi X}
=
K_{{\rm J}/\psi}\sum_{a,b}
\int d q^2 \left ( \frac{
\hat \sigma_{ab\to c \bar c}(Q^2)}{Q^2}\right )
\nonumber\\
&& \times
\int dx_F \phi_{a/A}(x_a)\phi_{b/B}(x_b)
\frac{x_a x_b}{x_a+x_b}F_{c\bar c\to {\rm J}/\psi}(q^2),
\nonumber \\
\label{eq:xsec}
\end{eqnarray}
where $\sum_{a,b}$ runs over all parton flavors,
 $Q^2=q^2+4 m_c^2$, $\phi_{a/A}(x_a)$ is the distribution function
of parton $a$ in hadron $A$, and
$x_F=x_a-x_b$ and $x_a x_b=Q^2/s$.
$\hat \sigma$ is the parton cross section 
and $K_{{\rm J}/\psi}$ is a phenomenological constant.
The factor
$F_{c\bar c\to {\rm J}/\psi}(q^2)$, which describes 
the transition probability for
the $c\bar c$ state of the relative momentum $q^2$ to evolve into
a physical J/$\psi$, is parametrized as
\begin{equation}
F_{c\bar c\to {\rm J}/\psi}(q^2)
=N_{{\rm J}/\psi}\ \theta(q^2)\ \theta(4{m'}^2-4m_c^2-q^2)
\left (1-\frac{q^2}{4{m'}^2-4m_c^2}
\right )^{\alpha_F} .
\label{eq:powerfrag}
\end{equation}
This form includes the effect of the open charm threshold
at $4{m'}^2$ and simulates the gluon radiation effect with the parameter 
$\alpha_F>0$ which puts the larger weight to the smaller $q^2$.

The nuclear effect is taken into account 
as the coherent multiple scattering of the pair,
which
at the leading of the nuclear enhancement
 may result in shifting of the relative momentum 
in the transition probability,~\cite{QVZ02,RJF02}
\begin{equation}
F_{c\bar c\to {\rm J}/\psi}(\bar q^2)
=F_{c\bar c\to {\rm J}/\psi}(q^2+\varepsilon^2 L),
\label{eq:shift}
\end{equation}
where $L$ is the effective length of the nuclear medium
in the $AB$ collisions.
The model assumption here is the separation of the multiple scattering
regime from the later formation stage of the physical resonance.
We note here that for a large enough $L$ such that
$\bar q^2>4{m'}^2-4m_c^2$
the transition probability
essentially vanishes due to the existence of the open charm
threshold~(\ref{eq:powerfrag}).
This apparently gives a 
stronger suppression than the exponential one
which follows in the Glauber model.

The vanishing of the probability for the large $L$
may be altered due to the subleading nuclear effects, by which we
expect a diffusion of the momentum distribution.\cite{HF03}
We model the momentum diffusion here 
 by modifying the transition probability
$F_{c\bar c\to {\rm J}/\psi}$.
The $c\bar c$ pair with relative momentum $q$
produced in a hard parton collision, will change its momentum
to $q'$ after the random multiple scattering, and then transforms into
the J/$\psi$ with the probability $F_{c\bar c\to {\rm J}/\psi}({q'}^2)$.
After many scatterings, this classical, elementary diffusion process 
of the momentum results 
in  the Gaussian distribution around the initial value $q$ 
with the variance $\varepsilon ^2 L$.
For demonstration, we replace the transition probability 
by
\begin{equation}
\bar F_{c\bar c\to {\rm J}/\psi}(q^2)
 \equiv \frac{1}{(2\pi \varepsilon^2 L)^{3/2}}
\int d^3 q' e^{- \frac{(q'-q)^2}{2\varepsilon^2 L}}
F_{c\bar c\to {\rm J}/\psi}({q'}^2).
\label{eq:smear}
\end{equation}
Note that
 $\bar F_{c\bar c\to {\rm J}/\psi}(q^2)$
never vanishes for any $q$ although
the average momentum of the pair increases
as $\langle {q'}^2\rangle=q^2+3 \varepsilon^2 L$
 after the multiple scattering.
Importantly, 
the transition probability behaves
more moderately 
as $\bar F_{c\bar c\to {\rm J}/\psi}(q^2) \propto  L^{-3/2}$
for the asymptotically large $L$, 
which stems from the depletion of the normalization factor.

In Fig.~\ref{fig1}
we show our result on the J/$\psi$ suppression 
calculated using the formula (\ref{eq:xsec})
with the smeared probability (\ref{eq:smear})
with
$\varepsilon^2=0.185$ GeV$^2$/fm.
Other parameters are fixed to the same as~\citen{QVZ02}:
$\alpha_F=1$ and $f_{{\rm J}/\psi}\equiv 
K_{{\rm J}/\psi} N_{{\rm J}/\psi}=0.485$.
Our model reasonably fits the data in the p$A$ and $AB$
collisions taken from~\citen{NA50}
except the Pb-Pb point. The curve bends upward slightly
in the semi-log plot.
The original QVZ model (\ref{eq:shift})
with $\varepsilon^2=0.25$ GeV$^2$/fm 
(dashed line) can explain  all the data points in Fig.~\ref{fig1}.
The downward bending of QVZ model 
is the result of the existence of the open charm 
threshold in~(\ref{eq:powerfrag}) and the uniform momentum
shift~(\ref{eq:shift}).

This simple analysis indicates the importance of the
subleading effect or the momentum diffusion to
the $L$-dependence of the J/$\psi$ formation.

\section{High-energy $Q\bar Q$ state passing through random gauge fields}

The coherent multiple scattering of the pair becomes more important at the
collider energies, where two colliding nuclei are Lorentz-contracted to
thin slabs and therefore the interaction of the $c \bar c$ pair
with the valence nucleons should be treated as a simultaneous action.

Let us study a simpler problem
of the high-energy heavy $Q\bar Q$ pair passing through a nucleus,
which is modeled as a random filed source,
and demonstrate that the penetration probability of the singlet bound state
decays non-exponentially.\cite{FM02}
Traveling through the target nucleus, the quark-anti-quark
pair acquires the
eikonal phase,
\begin{eqnarray}
&&U(\bx,\by;A)
=W(\bx;A)W^\dagger(\by;A)  ,
\label{eq:eik}
\end{eqnarray}
accumulated along the path with the transverse
positions, $\bx$ and $\by$, with $W(\bx;A)=
{\cal P}\exp[ig\int_{-\infty}^{\infty} d x^+ A^{a-}(\bx,x^+)t^a]$.
This is a matrix in the color space
and ${\cal P}$ indicates 
\begin{wrapfigure}[7]{h}{0.38\textwidth}
\epsfxsize=0.36\textwidth\epsffile{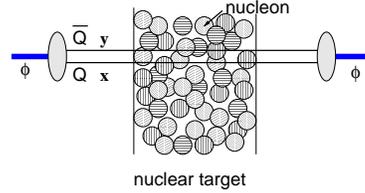}
\caption{High-energy $Q\bar Q$ pair 
passing through a nuclear target.\label{target}}
\end{wrapfigure}
the path ordering
of the product. 
In the definite 
color basis
they are written as
\begin{eqnarray}
U_{\rm ss}(\bx,\by) &=&
\frac{1}{N}
\Tr(W(\bx)W^\dagger(\by)),
\nonumber \\
U^a_{\rm as}(\bx,\by) &=&
\sqrt{\frac{2}{N}}
\Tr(W(\bx) W^\dagger(\by) t^a),
\nonumber \\
U^{ba}_{\rm aa}(\bx,\by) &=&
2\Tr( t^b W(\bx) t^a W^\dagger(\by)).
\label{eq:coloramp}
\end{eqnarray}
Here the color transparency is manifest as $U_{\rm ss}(\bx,\bx) =1$
and $U^a_{\rm as}(\bx,\bx) =0$.
The survival probability of the color singlet $Q\bar Q$ state 
$\varphi(\br,z)$ ($\br=\bx-\by$ and $z$ is the longitudinal
momentum fraction) is written as
$S \sim |\int d^2 \br dz \varphi(\br,z) U_{\rm ss} \varphi(\br,z) |^2$.

As we are interested only in the $Q\bar Q$ state, we take a closure
of the final target states and average the probability over the
initial target state. This is performed by dividing the
path into small zones and averaging over the
random gauge configurations in Eq.~(\ref{eq:eik}),\cite{FM02}
which results in a kernel 
$\overline{U(\bx,\by)U^\dagger(\bar \bx,\bar \by)}
\equiv K(\br, \bar \br, \bR-\bar \bR)$
with $\bR=\frac{1}{2}(\bx+\by)$ and $\bar \bx$ the coordinate
in the conjugate amplitude. The $\br$ ($\bR$) dependence 
of $K$ gives rise to the relative (total) momentum diffusion
of the $Q\bar Q$ state
caused by this interaction with the random gauge fields.

\begin{wrapfigure}{h}{0.38\textwidth}
\epsfxsize=0.37\textwidth\epsffile{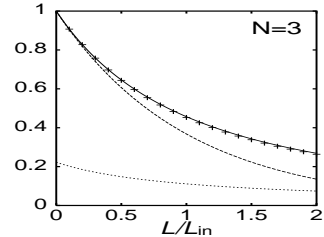}
\caption{Survival probability of the singlet Gaussian state
traversing SU(3) random fields [cross] is
compared with Eq.~(\ref{eq:analform}) [solid] 
and the exponential form [dashed].
The first term of (\ref{eq:analform}) is shown by a dotted line.
\label{survfig}}
\end{wrapfigure}
In the thin and thick target limits, one finds the asymptotic
behavior
of the survival probability,
$S = \int d^2 \br d^2 \bar \br \rho_0(\br) \rho_0(\bar \br)
K(\br,\bar\br,\bzero)$ with $\rho_0(\br)=\int dz \varphi^2(\br,z)$,
as
\begin{equation}
S \sim 1-\frac{L}{L_{\rm in}} \ (L\to 0), \quad
S \propto \frac{1}{N^2}\frac{L_{\rm in}}{L} \ (L\to \infty).
\label{eq:asympt}
\end{equation}
The first relation defines $L_{\rm in}$. The power-law suppression
$L^{-1}$ is the result of the 2-dimensional diffusion of the relative momentum,
which corresponds to $L^{-3/2}$  in \S2 where we assumed the 3-dimensional
random walk instead. The $1/N^2$ is the result of the color diffusion.

For concreteness we calculated the survival probability of a 
small Gaussian state as shown in Fig.~\ref{survfig}.
Furthermore we speculate an analytic approximation
which is compatible with the asymptotic behavior (\ref{eq:asympt}):
\begin{eqnarray}
S(L/L_{\rm in})= \frac{2}{N^2} \frac{1}{1+\frac{L}{L_{\rm in}}}
+\frac{N^2-2}{N^2} \frac{1}{(1+\frac{1}{2}\frac{L}{L_{\rm in}})^2}.
\label{eq:analform}
\end{eqnarray}
This formula is found to fit the numerical result quite well, and 
clearly demonstrates the non-exponential behavior of the survival
probability.

\section{Concluding Remarks}

It is a theoretical challenge to predict what will happen at 
BNL-RHIC before the new data show 
up.\cite{QM02,Phenix}
Our elementary demonstration is a example which indicates that the 
J/$\psi$ production at BNL-RHIC energy 
will be different from those observed at CERN-SPS energy.
At the same time
one should  think  over again the CERN-SPS data and the available
p$A$ data to construct a unified picture of the quarkonium production
in the nucleus-nucleus collisions.  
Such steady efforts will make the study 
with BNL-RHIC and CERN-LHC successful,
together with other various observables.

\section*{Acknowledgements}
The author is indebted to D.~Kharzeev and T.~Matsui for their
valuable suggestions and the fruitful collaborations on this subject.
This work is supported in part by  the Grants-in-Aid for Scientific Research 
of Monbu-kagaku-sho (No.\ 13440067).


\vspace{-0.2mm}

\begin{thebibliography}{99}
%%%%%%%%%%%%%%%%%%%%%%%%%%%%%%%%%%%%%%%%%%%%%%%%%%%%%%%%%%%%%
% Some macros are available for the bibliography:
%   o for general use
%      \JL : general journals          \andvol : Vol (Year) Page
%   o for individual journal 
%      \PR  : Phys. Rev.               \PRL : Phys. Rev. Lett.
%      \NP  : Nucl. Phys.              \PL  : Phys. Lett.
%      \JMP : J. Math. Phys.           \CMP : Commun. Math. Phys.
%      \PTP : Prog. Theor. Phys.       \JPSJ: J. Phys. Soc. Jpn.
%      \JP  : J. of Phys.              \NC  : Nouvo Cim.
%      \IJMP: Int. J. Mod. Phys.       \ANN : Ann. of Phys.
% Usage:
%   \PR{D45,1990,345}            ==> Phys.~Rev.\ {\bf D45} (1990), 345
%   \JL{Phys.~Lett.,A30,1981,56} ==> Phys.~Lett.\ {\bf A30} (1981), 56
%   \andvol{B123,1995,1020}      ==> {\bf B123} (1995), 1020
%%%%%%%%%%%%%%%%%%%%%%%%%%%%%%%%%%%%%%%%%%%%%%%%%%%%%%%%%%%%%
\bibitem{QM01} T.J.~Hallman {\em et al.} ed.,
               Proc.~of {\em the 15th International Conference on 
               Ultra-Relativistic Nucleus-Nucleus Collisions}, 
               Nucl.~Phys.~A {\bf 698} (2002), 1.
\bibitem{QM02} Proc.~of {\em the 16th International Conference on
               Ultra-Relativistic Nucleus-Nucleus Collisions}, 
               Nucl.~Phys.~A, to be published.
\bibitem{matsui} T.~Matsui, in these proceedings.
\bibitem{Phenix} H.~Hamagaki, in these proceedings.
\bibitem{MS86} T.\ Matsui and H.\ Satz, Phys.\ Lett.\ B {\bf 178} (1986), 416.
\bibitem{NA38} C.~Baglin, {\em et al.} (NA38 collab.), Phys.~Lett.~B 
          {\bf 220} (1989), 471; B {\bf 255} (1991), 459.
\bibitem{GH92} C.~Gerschel and J.~H\"ufner, Z.~Phys.~C {\bf 56} (1992), 71.
\bibitem{NA38plb466} M. C. Abreu, {\em et al.} (NA38 collab.), Phys.~Lett.~B
          {\bf 466} (1999), 408, and references therein. %pA and AB J/psi
\bibitem{KLNS97} D.~Kharzeev, C.~Louren\c{c}o, M.~Nardi and H.~Satz, 
                 Z.~Phys. C {\bf 74}  (1997), 307.
\bibitem{KS96} D.~Kharzeev and H.~Satz, Phys.~Lett.~B {\bf 366} (1996), 316. 
               %color octet model
\bibitem{NA50} M.C.~Abreu, {\em et al.} (NA50 collab.)  Phys.~Lett.~B
{\bf 410} (1997), 337;  B {\bf 477} (2000), 28. %anomalous J/psi
\bibitem{BDO00} E.g., J.P.~Blaizot, P.M.~Dinh and J.Y.~Ollitrault,
                Phys.~Rev.~Lett.\ {\bf 85} (2000), 4012.
\bibitem{CFK00} E.g., A.~Capella, E.G.~Ferreiro and A.B.~Kaidalov, 
                Phys.~Rev.~Lett.\ {\bf 85} (2000), 2080.
\bibitem{cross} K.~Martins, D.~Blaschke and E.~Quack, 
          Phys.~Rev.~C {\bf 51} (1995),2723-2738.
          S.G.~Matinyan and B.~M\"uller, Phys.~Rev.~C {\bf 58} (1998), 2994.
          D.~Kharzeev and H.~Satz, Phys.~Lett.~B {\bf 334} (1994), 155.
          H.~Fujii and D.~Kharzeev, Phys.~Rev.~D {\bf 60} (1999), 114039;
          Nucl.~Phys.~A {\bf 661} (2000), 542c.
\bibitem{STST00} A.~Sibirtsev, K.~Tsushima, K.~Saito, A.W.~Thomas,
               Phys.\ Lett.\ B {\bf 484} (2000), 23.
\bibitem{AH00} A.\ Hayashigaki, Phys.\ Lett.\ B {\bf 487} (2000), 96.
\bibitem{coalesce}  P.~Braun-Munzinger and J.~Stachel, 
                   Phys.~Lett.~B{\bf 490} (2000), 196.\\
                    R.L.~Thews, M.~Schroedter and J.~Rafelski, 
                   Phys.~Rev.~ C {\bf 63} (2001), 054905.
\bibitem{QVZ02} J.W.\ Qiu, J.P.\ Vary and X.F.\ Zhang, Phys.\ Rev.\ Lett.\
{\bf 88} (2002), 232301.
\bibitem{RJF02} F.J.~Fries, hep-ph/0209275.
\bibitem{HF03} H.~Fujii, Phys.~Rev.~ C {\bf 67} (2003), 031901(R).
\bibitem{FM02} H.~Fujii and T.~Matsui, Phys.~Lett.~B {\bf  545} (2002), 82;
               Nucl.~Phys.~A {\bf 709} (2002), 236.
\end{thebibliography}
\end{document}